\documentclass[]{aa}

\def\rmit#1{{\it #1}}              

\def\eg{\rmit{e.g.}}

\usepackage{booktabs}
\usepackage{graphicx}
\usepackage{multirow}
\usepackage{color}
\usepackage[flushleft]{threeparttable}

\usepackage{array}
\newcolumntype{?}{@{\vrule width 2pt}}

\usepackage{hyperref}
\hypersetup{
    final=true,
    pageanchor=true,
    colorlinks=true,
    breaklinks=true,
    linkcolor=blue,
    citecolor=blue,
    urlcolor=blue,
    pdfpagemode=UseNone,
    pdftitle={},
    pdfauthor={T. Felipe et al.},
    pdfsubject={Solar Physics},
    pdfkeywords={Methods: observational -- Sun: photosphere -- Sun: chromosphere -- Sun: oscillations  -- sunspots -- Techniques: polarimetric}}

\titlerunning{}   

\begin{document}



\title{Origin of the chromospheric three-minute oscillations in sunspot umbrae}

\author{T. Felipe\inst{\ref{inst1},\ref{inst2}}
}


\institute{Instituto de Astrof\'{\i}sica de Canarias, 38205, C/ V\'{\i}a L{\'a}ctea, s/n, La Laguna, Tenerife, Spain\label{inst1}
\and 
Departamento de Astrof\'{\i}sica, Universidad de La Laguna, 38205, La Laguna, Tenerife, Spain\label{inst2} 
}

\abstract
{Sunspot umbrae show a change in the dominant period of their oscillations from five minutes (3.3 mHz) in the photosphere to three minutes (5.5 mHz) in the chromosphere.} 
{In this paper, we explore the two most popular models proposed to explain the three-minute oscillations: the chromospheric acoustic resonator and the propagation of waves with frequency above the cutoff value directly from lower layers.}
{We employ numerical simulations of wave propagation from the solar interior to the corona. Waves are driven by a piston at the bottom boundary. We have performed a parametric study of the measured chromospheric power spectra in a large number of numerical simulations with differences in the driving method, the height of the transition region (or absence of transition region), the strength of the vertical magnetic field, and the value of the radiative cooling time. }
{We find that both mechanisms require the presence of waves with period in the three-minute band at the photosphere. These waves propagate upward and their amplitude increases due to the drop of the density. Their amplification is stronger than that of evanescent low-frequency waves. This effect is enough to explain the dominant period observed in chromospheric spectral lines. However, waves are partially trapped between the photosphere and the transition region, forming an acoustic resonator. This chromospheric resonant cavity strongly enhances the power in the three-minute band.}
{The chromospheric acoustic resonator model and the propagation of waves in the three-minute band directly from the photosphere can explain the observed chromospheric three-minute oscillations. They are both important in different scenarios. Resonances are produced by waves trapped between the temperature minimum and the transition region. Strong magnetic fields and radiative losses remove energy from the waves inside the cavity, resulting in weaker amplitude resonances.}


\maketitle


\section{Introduction}

Solar magnetically active regions exhibit many oscillatory phenomena. Their study is a relevant topic in astrophysics, since they can be used to infer the properties of the media where they propagate using seismology techniques \citep{deMoortel+Nakariakov2012}, whereas they have also been proposed as one of the candidates to explain chromospheric and coronal heating \citep{Alfven1947}. Since the first detection of waves in sunspots \citep{Beckers+Tallant1969}, many oscillatory phenomena have been reported in various magnetic structures, from the photosphere to the outer layers. In the case of sunspots, photospheric umbral oscillations show a period of five minutes \citep{Bhatnagar+etal1972}, while at the umbral chromosphere their period is reduced to three minutes \citep{Beckers+Schultz1972} and they can form shocks \citep{Lites1986, Centeno+etal2006}. All these manifestations are associated to a global process of wave propagation in sunspots \citep[see][for a review]{Khomenko+Collados2015}. In fact, several authors have provided evidence of wave propagation from the photosphere to the corona \citep{Reznikova+etal2012, KrishnaPrasad+etal2015, Zhao+etal2016}.

One of the basic observational findings in umbral atmospheres is the reduction of the period of the oscillations with increasing height, from five minutes in the photosphere to three minutes in the chromosphere \citep[\eg,][]{Lites1992}.  Three mechanisms have been proposed to explain the origin of the three-minute oscillations: propagation of waves with frequency above the cutoff directly from the photosphere, a chromospheric resonant cavity, and their formation in the wake of propagating shocks. The later is found in non-linear one dimensional simulations \citep{Hansteen1997, Bard+Carlsson2010}, although \citet{Ulmschneider+etal2005} concluded that the shock mergins leading to this effect are unrealistic. In this work we focus in the two other models. \citet{Centeno+etal2006} fitted the phase spectrum between the Doppler velocity from photospheric and chromospheric lines with a model of linear vertical propagation of slow magnetoacoustic waves in a stratified atmosphere with radiative losses. They conclude that waves with frequency above the cutoff value can propagate upwards, and due to the rapid drop of the density their amplification is stronger than that of evanescent low-frequency waves (with frequency below the cutoff). This physical effect was previously modeled by \citet{Fleck+Schmitz1991}. In their numerical simulations, waves were driven at the bottom boundary by a long-period sinusoidal function in the vertical velocity. This driver also generates high-frequency waves, which propagate upward and dominate the power spectra at the chromosphere. Later, \citet{Felipe+etal2010a}  modeled this mechanism by explicitly using a driver that introduced waves in the three-minute band below the photosphere, and \citet{Felipe+etal2011} performed observationally driven simulations that reproduce the observed chromospheric three-minute oscillations by means of high-frequency waves propagating from deeper layers. 

The second mechanism explored in this work is that of a chromospheric acoustic resonator \citep{Zhugzhda+Locans1981, Zhugzhda+etal1983,Gurman+Leibacher1984}. Slow magnetoacoustic waves are trapped between the height where the cutoff frequency is below the wave frequency (around the temperature minimum) and the steep temperature gradients at the transition region. These boundaries are permeable and form a resonant cavity. \citet{Zhugzhda2008} analyzed the effect of this resonator taking into account the changes in the temperature with height. This mechanism has also been studied through numerical simulations using the ideal magnetohydrodynamic (MHD) approximation \citep{Botha+etal2011, Snow+etal2015} and using a two-fluids model \citep{Wojcik+etal2018}. They support the existence of an acoustic resonator that generate the observed chromospheric oscillations in the three-minute band. It must be noted that the {\bf resonator} does not prevent the effect of the cutoff frequency.

Observational works have reported the presence of three-minute oscillations in the umbral photosphere \citep[\eg,][]{Lites+Thomas1985, Abdelatif+etal1986,Centeno+etal2006, Felipe+etal2010b}. \citet{Chae+etal2017} suggested that the three-minute oscillations are generated by magnetoconvection occurring in the umbra (in light bridges and umbral dots), and reported photospheric three-minute oscillations with energy flux large enough to explain the chromospheric oscillations as upwardly propagating slow magnetoacoustic waves. \citet{Stangalini+etal2012} found a three-minute wave enhancement in the photosphere of a pore, whose amplitude increases with the magnetic field strength.

In this paper, we numerically investigate wave propagation in sunspot umbrae. We have performed a set of numerical experiments that explore two of the mechanisms discussed in previous paragraphs: the chromospheric acoustic resonator and the propagation of waves in the three-minute band directly from the photosphere. We evaluate the effect of various physical ingredients on the results. In Sect. \ref{sect:simulations} we describe the numerical simulations and the results are presented in Sect. \ref{sect:cavity}. Sections \ref{sect:discussion} and \ref{sect:conclusions} show the discussion and conclusions, respectively.

\section{Numerical simulations}
\label{sect:simulations}

Wave propagation from the solar interior to the corona has been computed using the code MANCHA \citep{Khomenko+Collados2006, Felipe+etal2010a}. In this paper, we have restricted the use of the code to solve the ideal non-linear MHD equations in a one dimensional atmosphere, but keeping the three dimensional components of the vectors. The equilibrium state is explicitly removed from the equations. The code solves the following system of equations for perturbations:

\begin{equation}
\frac{\partial\rho_1}{\partial
t}+\nabla\Big[(\rho_0+\rho_1){\bf v_1}\Big]= 0 \,, \label{eq:cont1}
\end{equation}

\begin{eqnarray}
\lefteqn{\frac{\partial [(\rho_0+\rho_1){\bf v}_1]}{\partial t}+\nabla \cdot\Big [(\rho_0+\rho_1){\bf
v}_1{\bf v}_1+\Big(p_1+}\nonumber\\
&&+\frac{{\bf B}_1^2}{2\mu_0}+\frac{{\bf B}_1{\bf
B}_0}{\mu_0}\Big){\bf I}- \frac{1}{\mu_0}({\bf B}_0{\bf B}_1-{\bf
B}_1{\bf
B}_0-{\bf B}_1{\bf B}_1)\Big ]=\nonumber\\
&&=\rho_1 {\bf g}\,, \label{eq:mom1}
\end{eqnarray}

\begin{eqnarray}
\lefteqn{\frac{\partial e_1}{\partial t}+\nabla\cdot\Big[{\bf
v_1}\Big((e_0+e_1)+(p_0+p_1)+}\nonumber\\
&&+\frac{|{\bf B}_0+{\bf B}_1|^2}{2 \mu_0}\Big)- \frac{1}{\mu_0}({\bf B}_0+{\bf B}_1)\Big({\bf v}_1 \cdot ({\bf B}_0 + {\bf B}_1)\Big)\Big ]=\nonumber\\
&&=(\rho_0+\rho_1) ({\bf g\cdot v_1})+Q_{\rm rad}\,,
\label{eq:ene1}
\end{eqnarray}

\begin{equation}
\frac{\partial {\bf B}_1}{\partial t}=\nabla\times [{\bf v_1} \times ({\bf B}_0+{\bf B}_1)] \,,
\label{eq:ind1}
\end{equation}

\noindent where $\rho$ is the density, $p$ is the gas pressure, ${\bf v}$ is the velocity, ${\bf
B}$ is the magnetic field, {\bf I} is the identity tensor and $e$ is the total energy per unit volume, given by

\begin{equation}
e=\frac{1}{2}\rho v^2 + \frac{p}{\gamma-1} + \frac{B^2}{2\mu_0} \,.
\end{equation}

\noindent Subindex 0 indicates the background values, corresponding to an atmosphere in equilibrium, while variables with subindex 1 are the perturbed quantities. The gravitational acceleration and the magnetic permeability are given by ${\bf g}$ and $\mu_{\rm 0}$, respectively. The term $Q_{\rm rad}$ represent the radiative losses (see Sect. \ref{sect:rad_trans}). Artificial diffusion terms were added following \citet{Vogler+etal2005}.

\begin{figure}[!ht] 
 \centering
 \includegraphics[width=9cm]{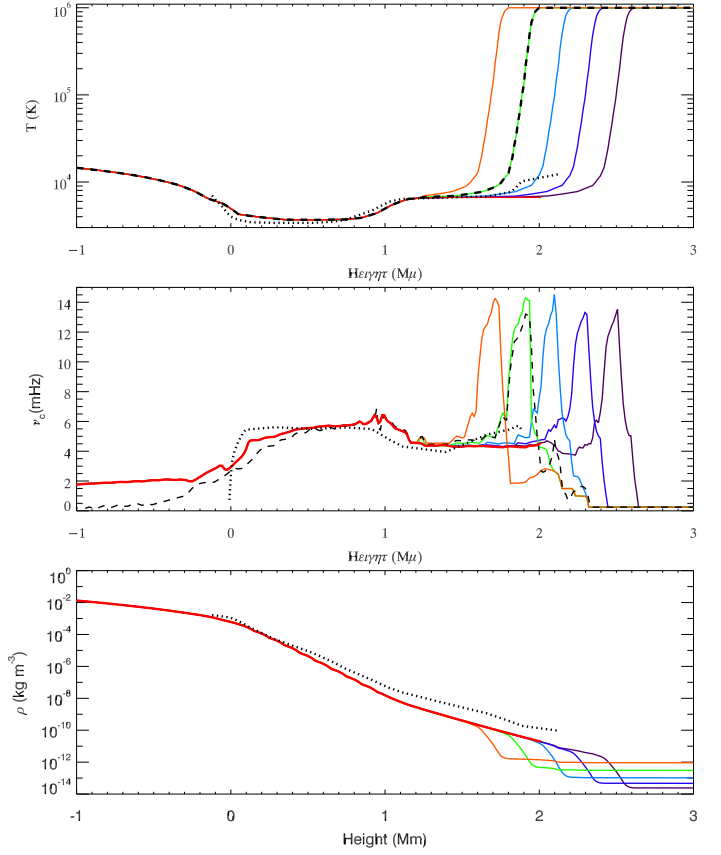}
  \caption{Temperature (top panel), acoustic cutoff frequency (middle panel), and density (bottom panel) of the sunspot models. They differ in the height of the base of the corona, which is located at 1.8 Mm (orange), 2.0 Mm (green), 2.2 Mm (light blue), 2.4 Mm (dark blue, and 2.6 Mm (violet). The black dashed line correspond to the magnetoacoustic cutoff of the umbral model with the base of the corona at 2.0 Mm and a vertical magnetic field of 2500 G. The red line is the model without corona. The dotted black line corresponds to the sunspot model M from \citet{Maltby+etal1986}.}      
  \label{fig:models}
\end{figure}

Umbral models have been constructed by combining the convectively stabilized CSM\_B interior model \citep{Schunker+etal2011} with the photospheric and chromospheric atmosphere obtained from \citet{Avrett1981} umbral model. A one million Kelvin isothermal corona was added following \citet{Santamaria+etal2015}, with a sharp temperature increase from the chromospheric value to the coronal temperature in a 200 km thick transition region. Various umbral models have been constructed. They differ in the height of the base of the isothermal corona, which ranges from $z=1.8$ Mm to $z=2.6$ Mm. The photosphere is located at the height $z=0$, where the optical depth at 500 nm is unity. Figure \ref{fig:models} shows the temperature, cutoff frequency, and density profiles of those models. The sunspot model M from \citet{Maltby+etal1986} is also plotted as a reference. In the top panel, the color lines indicate the acoustic cutoff of various umbral models (with differences in the height of the transition region) obtained from the expression:

\begin{equation}
\omega_{\rm c}=\frac{c_{\rm s}}{2H}\Big (1-2\frac{dH}{dz}\Big )^{1/2},
\label{eq:cutoff}
\end{equation}

\noindent where $H=c_s^2/(\gamma g)$ is the pressure scale height, $\gamma$ is the coefficient of specific heats, $c_{\rm s}$ is the sound speed, and $g$ is the gravity. For the model with the base of the corona at $z=2.0$ Mm, it is also shown the cutoff frequency for magnetoacoustic waves in an atmosphere permeated by an uniform vertical magnetic field of $B=2500$ G, as derived by  \citet{Roberts2006}:

\begin{equation}
\omega_{\rm c}^{\rm B}=c_{\rm t}\Big [\frac{1}{4H^2} \Big (\frac{c_{\rm t}}{c_{\rm s}}\Big )^4-\frac{1}{2}\gamma g\frac{\partial}{\partial z}\Big (\frac{c_{\rm t}^2}{c_{\rm s}^4}\Big )+\frac{1}{v_{\rm A}^2}\Big (N^2+\frac{g}{H}\frac{c_{\rm t}^2}{c_{\rm s}^2}\Big )\Big ]^{1/2}, 
\label{eq:cutoff_B}
\end{equation} 

\noindent where 

\begin{equation}
N^2=-g\Big (\frac{1}{\rho}\frac{\partial\rho}{\partial z}-\frac{1}{\gamma p}\frac{\partial p}{\partial z}\Big )
\label{eq:brunt}
\end{equation}

\noindent is the squared Brunt-V\"ais\"al\"a frequency, $v_A=B/(4\pi\rho)^{1/2}$ is the Alfv\'en speed, and $c_{\rm t}=c_{\rm s}v_{\rm A}/(c_{\rm s}^2+v_{\rm A}^2)^{1/2}$ is the cusp speed.

We have performed several numerical experiments. In some of them, we aim to explore the effect of a chromospheric cavity formed between the temperature minimum and the transition region. In those cases, the top of the computational domain is placed at $z=5.4$ Mm. At the top boundary we have imposed a Perfect Matched Layer \citep{Berenger1994}, which damps the waves and avoids the presence of spurious reflections. In the PML domain, the MHD equations (Eqs. \ref{eq:cont1}-\ref{eq:ind1}) are split into a set of three coupled one dimensional equations, and a damping coefficient is added to each equation only near the boundary of the corresponding direction. Numerical tests with differences in the height and strength of the PML top boundary has been performed, and no significant differences have been found in the results. For further details about the PML, we address the reader to \citet{Berenger1994}, \citet{Hu1996}, and \citet{Felipe+etal2010a}.

In order to study wave propagation in a model with no chromospheric cavity, we have used an umbral atmosphere with constant chromospheric temperature above $z=1.4$ Mm and with the top boundary located at $z=2$ Mm (red line in Fig. \ref{fig:models}). In all simulations, the bottom boundary is placed at $z=-5.2$ Mm. The code employs a variable spatial resolution. The grid is stretched in the vertical coordinate according to the local sound speed. A coarser resolution is used in regions with high sound speed and a finer resolution is regions with low sound speed. The travel time of acoustic waves between adjacent cells is the same in all the domain, except at layers above the temperature minimum (including chromosphere, transition region, and corona), where we have forced the finest resolution ($\Delta z=12.4$ km). The lowest resolution ($\Delta z=49.5$ km) is set at the bottom boundary.

Waves are driven by specifying a perturbation as a function of time in a few grid points at the bottom boundary. We have employed two different drivers. First, we have implemented a monochromatic acoustic-gravity wave of a selected period. We have imposed perturbations in pressure, density, and velocity obtained from analytical calculations \citep{Mihalas+Mihalas1984}. These perturbations do not take into account the temperature gradient and magnetic field (whose effect is negligible at this depth). According to \citet{Khomenko+Cally2012}, the analytical expressions for the perturbations in velocity, pressure, and density are given by

\begin{center}
\begin{equation}
\label{eq:mono_vz} 
v_{z1}=V\exp\Big(\frac{z}{2H}+k_{zi}z\Big )\sin (\omega t-k_{zr}z-k_xx)
\end{equation}
\end{center}

\begin{center}
\begin{equation}
\label{eq:mono_p} 
\frac{p_{1}}{p_0}=V|P|\exp\Big(\frac{z}{2H}+k_{zi}z\Big )\sin (\omega t-k_{zr}z-k_xx+\phi_P)
\end{equation}
\end{center}

\begin{center}
\begin{equation}
\label{eq:mono_rho} 
\frac{\rho_{1}}{\rho_0}=V|R|\exp\Big(\frac{z}{2H}+k_{zi}z\Big )\sin (\omega t-k_{zr}z-k_xx+\phi_R)
\end{equation}
\end{center}

\begin{center}
\begin{equation}
\label{eq:mono_vx} 
v_{x1}=V|U|\exp\Big(\frac{z}{2H}+k_{zi}z\Big )\sin (\omega t-k_{zr}z-k_xx+\phi_U)
\end{equation}
\end{center}

\noindent where the amplitude of the wave is controlled by the parameter $V$. The relative phase shifts and amplitudes follow

\begin{center}
\begin{equation}
\label{eq:mono_ampl_P} 
|P|=\frac{\gamma\omega}{\omega^2-c_{\rm s}^2k_x^2}\sqrt{k_{zr}^2+\Big (k_{zi}+\frac{1}{2H}\frac{(\gamma-2)}{\gamma}\Big )^2}
\end{equation}
\end{center}

\begin{center}
\begin{equation}
\label{eq:mono_ampl_R} 
|R|=\frac{\omega}{\omega^2-c_{\rm s}^2k_x^2}\sqrt{k_{zr}^2+\Big (k_{zi}+\frac{(\gamma -1)}{\gamma H}\frac{c_{\rm s}^2k_x^2}{\omega^2}-\frac{1}{2H}\Big )^2}
\end{equation}
\end{center}

\begin{center}
\begin{equation}
\label{eq:mono_ampl_U} 
|U|=\frac{c_{\rm s}^2k_x}{\gamma\omega}|P|
\end{equation}
\end{center}

\begin{center}
\begin{equation}
\label{eq:mono_phase_PU} 
\phi_P=\phi_U=\arctan\Big (\frac{k_{zi}}{k_{zr}}+\frac{1}{2Hk_{zr}}\frac{(\gamma-2)}{\gamma}\Big )
\end{equation}
\end{center}

\begin{center}
\begin{equation}
\label{eq:mono_phase_R} 
\phi_R=\arctan\Big (\frac{k_{zi}}{k_{zr}}+\frac{(\gamma-1)}{\gamma Hk_{zr}}\frac{c_{\rm s}^2k_x^2}{\omega^2}-\frac{1}{2Hk_{zr}} \Big ).
\end{equation}
\end{center}

\noindent For a given frequency $\omega$, the vertical wavenumber $k_{z}$ is obtained from the dispersion relation for acoustic-gravity waves

\begin{center}
\begin{equation}
\label{eq:mono_phase_R} 
k_z=k_{zr}+ik_{zi}=\sqrt{\frac{(\omega^2-\omega_{\rm c}^2)}{c_{\rm s}^2}-\frac{k_x^2(\omega^2-N^2)}{\omega^2}},
\end{equation}
\end{center}

\noindent since we have imposed vertical wave propagation and, thus, the horizontal wavenumber $k_{x}$ has been set to $0$. All variables oscillate with a sinusoidal function during the whole duration of the simulation, whose period was set to $P=2\pi/\omega =300$ s.  

The second kind of driver is a Ricker wavelet introduced in the vertical velocity at the bottom boundary. Its temporal variation is given by       

\begin{center}
\begin{equation}
\label{eq:broadband} 
v_z(t)=A(1-2\tau_0^2)e^{-\tau_0^2}
\end{equation}
\end{center}

\noindent where $\tau_0 =0.5\omega t -\pi$ and $A$ is the amplitude. This driver excites a broadband wave spectrum with its central peak at frequency $\omega$, which was chosen to correspond to a period $P=300$ s. Drivers with this temporal evolution have been shown to excite a wave spectrum which resembles the solar observations \citep[\eg,][]{Parchevsky+etal2008, Felipe+etal2016a}.

In this paper we have performed a set of one dimensional numerical experiments of wave propagation in a sunspot umbra. The set up of the simulations includes differences in the background atmosphere, the location of the top boundary, the presence of magnetic field and radiative losses, and the amplitude and nature of the wave driving mechanism. These simulations allow us to explore different aspects about the origin of three-minute oscillations in umbral chromospheres. Tables \ref{table:cavity_monochromatic}-\ref{table:spiegel} show a summary of the simulations discussed in each section. Some of the details about the implementation of the numerical simulations are addressed in the corresponding section.

\section{Results}
\label{sect:cavity}

\subsection{Monochromatic driver}
\label{sect:cavity_monochromatic}

\begin{table*}
\begin{center}
\caption[]{\label{table:cavity_monochromatic}
          {Summary of the numerical simulations performed for Sect. \ref{table:cavity_monochromatic}.}}
\begin{tabular}{lcccccc}
\hline\noalign{\smallskip}
Simulation \#				& 	Driver 		&	Height Cor.	& B 		&  Ampl		&$\tau_{\rm R}$ 	 &Power peak 		\\
					& 	 		&	 (Mm)		& (G)		&  (m s$^{-1}$)	& (s)			 &(s)			\\
\hline\noalign{\smallskip}	
\ref*{sect:cavity_monochromatic}.1	& 	Monochromatic	& 	2.0		& 0		& 10$^{-3}$	&	$\infty$	&	300.0		\\	
\ref*{sect:cavity_monochromatic}.2	& 	Monochromatic	& 	-		& 0		&10$^{-3}$	&	$\infty$	&	300.0	\\

\hline

\end{tabular}

\begin{tablenotes}
\small
 \item {The first column indicates the number used to label the numerical simulation, the second column (Driver) is the kind of driver used at the bottom boundary (``Monochromatic'' for a monochromatic acoustic wave and ``Broadband'' for a broadband wavelet, in both cases with the power peak at a period of 300 s), the third column (Height Cor.) is the height of the bottom of the corona (a dash indicates that there is neither transition region nor corona), the fourth column (B) is the strength of the vertical magnetic field, the fifth column (Ampl) is the amplitude of the wave at the photosphere ($z=0$ Mm), the sixth column ($\tau_{\rm R}$) is the radiative cooling time, and the last column (Power peak) is the period of the maximum chromospheric power in the vertical velocity.}
\end{tablenotes} 
  
\end{center}
\end{table*}

In this section we present the results from the simulations shown in Table \ref{table:cavity_monochromatic}. In these numerical simulations we have introduced a sinusoidal acoustic wave with period of 300 s in the solar interior. The computational domain extends to the corona. Between the temperature minimum and the transition region a chromospheric cavity is formed, as shown by the local values of the acoustic cutoff frequency (green line from Fig. \ref{fig:models}). At around $z=1$ Mm the acoustic cutoff exhibits a local maximum around 6 mHz. This value is in agreement with the maximum cutoff determined observationally for a sunspot umbra by \citet{Felipe+etal2018b}. At the transition region (in this umbral model around $z=1.8$ Mm) the acoustic cutoff shows a prominent peak. Acoustic waves with frequency in the range $4.5-6$ mHz can propagate between $z\sim1$ Mm and the transition region, but they are trapped there. The amplitude of the wave has been chosen arbitrarily small in order to keep the simulation in the linear regime, and the magnetic field strength has been set to zero. The effects of magnetic field and realistic amplitudes will be analyzed later.

\begin{figure}[!ht] 
 \centering
 \includegraphics[width=9cm]{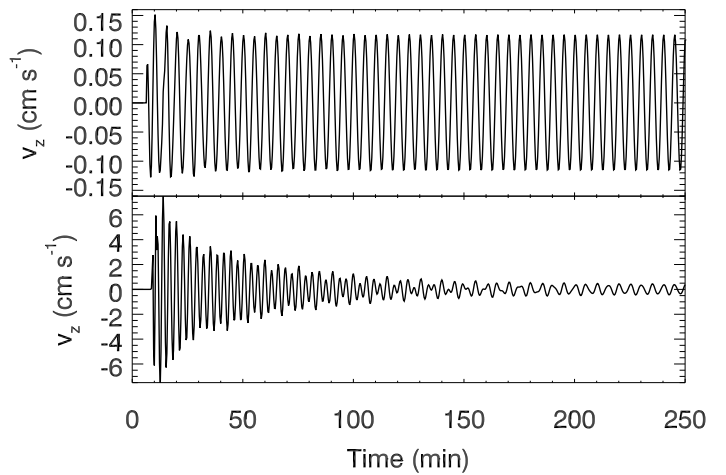}
  \caption{Temporal evolution of the vertical velocity at $z=0.3$ Mm (top panel) and at $z=1.4$ Mm (bottom panel) for Simulation \ref*{sect:cavity_monochromatic}.1.}      
  \label{fig:monochromatic_cavity_vz}
\end{figure}

Figure \ref{fig:monochromatic_cavity_vz} illustrates the temporal evolution of the vertical velocity at the photosphere (top panel) and the chromosphere (bottom panel). The photospheric velocity shows the regular pattern of a sinusoidal wave with a five minutes period and constant amplitude. This pattern takes place during the whole duration of the simulation (500 min, Fig. \ref{fig:monochromatic_cavity_vz} only shows the first half of the simulation), but the initial $\sim50$ min. The first wavefront driven at the bottom boundary $z=-5.2$ Mm needs approximately six minutes to reach the photosphere, and during this time there is no wave perturbation. The first photospheric wavefronts exhibit changes in their amplitude. The changes at the beginning of the simulation are more evident at the chromosphere (bottom panel of Fig. \ref{fig:monochromatic_cavity_vz}). The initial chromospheric wavefronts show high amplitude waves with a period around three minutes. The amplitude of these three-minute waves progressively decreases, until they completely vanish at $t\sim230$ min. After that time, the chromospheric velocity shows a steady sinusoidal wave with 5 min period. In the stationary stage of the simulation, an amplitude increase is found between the photosphere and chromosphere due to the drop of the density. This amplification is low because five-minute waves are evanescent (their frequency is below the acoustic cutoff).

\begin{figure}[!ht] 
 \centering
 \includegraphics[width=9cm]{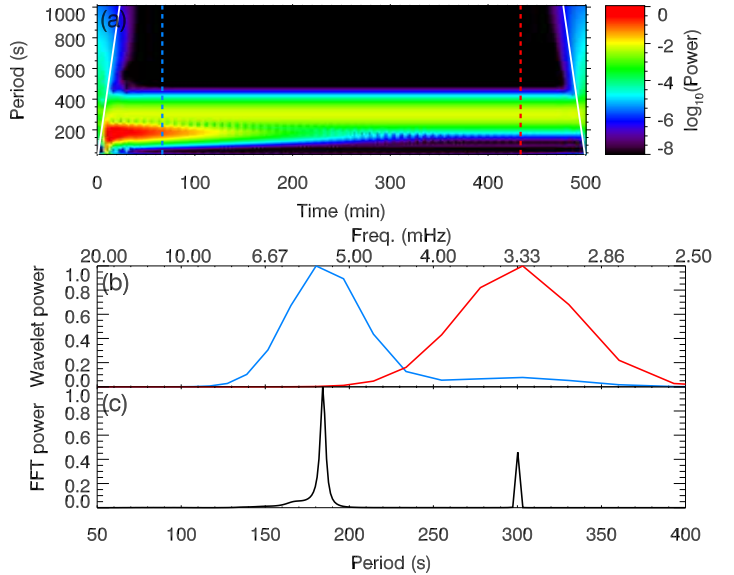}
  \caption{Top panel: Power as a function of period (vertical axis) and time (horizontal axis) obtained from the wavelet analysis of the vertical velocity at $z=1.4$ Mm from Simulation \ref*{sect:cavity_monochromatic}.1. The power is normalized and shown in a logarithmic scale. The results are reliable for the region within the two white lines. The vertical dashed lines indicate the time steps plotted in the middle panel. Middle panel: Normalized power at $t=67$ min (blue line) and at $t=430$ min (red line) obtained from the wavelet analysis. Bottom panel: Normalized power for the full temporal series (500 min) obtained from the FFT.}  
  \label{fig:monochromatic_cavity_wavelet}
\end{figure}

The change of the dominant period with time is clearly illustrated in the wavelet analysis \citep{Torrence+Compo1998} plotted in Fig. \ref{fig:monochromatic_cavity_wavelet}. The top panel shows the distribution of the power at the chromosphere as a function of time. Initially, strong power is concentrated in the three-minute band (blue line in the middle panel of Fig. \ref{fig:monochromatic_cavity_wavelet}). Oscillations in the five-minute band are present during the whole simulation (after the first wavefronts reach the chromosphere), and they become dominant when the three-minute oscillations fade away (red line in the middle panel of Fig. \ref{fig:monochromatic_cavity_wavelet}).

\begin{figure}[!ht] 
 \centering
 \includegraphics[width=9cm]{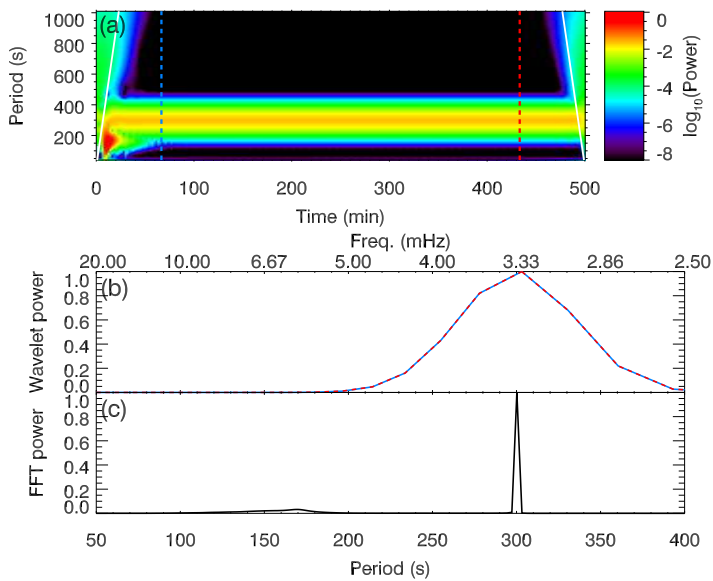}
  \caption{Same as Fig. \ref{fig:monochromatic_cavity_wavelet} but for Simulation \ref*{sect:cavity_monochromatic}.2. In this simulation there is no transition region.}      
  \label{fig:monochromatic_noTR_wavelet}
\end{figure}

Figure \ref{fig:monochromatic_noTR_wavelet} shows the results of the spectral analysis of Simulation \ref*{sect:cavity_monochromatic}.2. This simulation has the same set up as the previous case, except that this model does not include a transition region and corona. Similarly to the previous simulation, power in the three-minute band appears at the beginning of the computation, but it rapidly vanishes. The stationary state of the simulation is reached after just 30 min. During most of the simulation the chromospheric oscillations are dominated by a sharp five-minute peak. Even though we are introducing a monochromatic sinusoidal oscillation in the bottom boundary, the driven wave is not purely monochromatic during the first time steps. The atmosphere is initially at rest, and the wave is only present at a few grid points. Due to the finite extension of the wave pattern, other frequencies are inevitably excited, including high-frequency waves. These high-frequency waves propagate to higher layers and their amplitude increases due to the drop of the density. Above $z=0$, this increment is higher than that of the evanescent five-minute waves (see Fig. \ref{fig:models}), and they become dominant at the chromosphere. Not only waves with frequency above the local maximum in the acoustic cutoff frequency ($\sim6$ mHz) can reach the chromosphere. Waves with frequency between 4.5 and 6 mHz are evanescent in a small region, but they can leak to the chromosphere where they can freely propagate again. 

In Simulation \ref*{sect:cavity_monochromatic}.2 (without corona), the waves escape the computational domain through the top boundary, and a stationary simulation is soon obtained as the wave driven at the bottom boundary becomes purely monochromatic. However, in the simulation with transition region and corona (Simulation \ref*{sect:cavity_monochromatic}.1), high frequency waves are partially reflected at the transition region due to the strong temperature gradient. This reflected waves travel downward and reach the local maximum in the acoustic cutoff frequency (Fig. \ref{fig:models}), where those waves with frequency below 6 mHz are partially reflected again. This way, a resonance is formed in the chromospheric cavity. The power of waves with a period of 184.3 s is greatly enhanced, and they produce a sharp power peak, as seen in the power obtained from the fast Fourier transform (FFT) of the whole simulation (bottom panel of Fig. \ref{fig:monochromatic_cavity_wavelet}). Since the temperature gradients are not perfect reflecting surfaces, the three-minute waves progressively leak the cavity, until their power vanishes.     

In summary, the driver employed in these simulations provides a constant source of five-minute waves, whereas some small amount of energy goes to other frequencies during the initial time steps. Those frequencies above the cutoff value can reach the chromosphere and excite the resonant mode. As they leak the cavity, their amplitude decreases (see bottom panel of Fig. 2), and the simulation reaches a stationary state where only five-minute power is found.

\subsection{Monochromatic driver with realistic wave amplitudes}
\label{sect:amplitude}

\begin{table*}
\begin{center}
\caption[]{\label{table:amplitude}
          {Summary of the numerical simulations performed for Sect. \ref{sect:amplitude}.}}
\begin{tabular}{lccccccc}
\hline\noalign{\smallskip}
Simulation \#			& 	Driver 		&	Height Cor.	& B 		&  Ampl		&$\tau_{\rm R}$ 	 &Power peak 	.	\\
				& 	 		&	 (Mm)		& (G)		&  (m s$^{-1}$)	& (s)			 &(s)			\\
\hline\noalign{\smallskip}	

\ref*{sect:amplitude}.1		& 	Monochromatic	& 	2.0		& 0		&200		&	$\infty$	&	300.0	\\
\hline

\end{tabular}
   
\begin{tablenotes}
\small
 \item {Same description as Table \ref{table:cavity_monochromatic}.}
\end{tablenotes} 
  
\end{center}
\end{table*}

Observations show that vertical velocity oscillations in a sunspot umbra have an amplitude of a few hundred meters per second \citep[\eg,][]{Bhatnagar+etal1972, Soltau+etal1976, Lites+etal1982, Centeno+etal2006, Felipe+etal2010b}. In the previous section we shown that monochromatic five-minute waves with low amplitude (in the linear regime) cannot excite chromospheric three-minute oscillations through resonances in a chromospheric cavity. In this section we aim to explore this effect using realistic amplitudes.

The amplitude of the waves in the atmosphere is controlled through the amplitude imposed to the driver at the bottom boundary. We have chosen the appropriate value which produces a wave with 200 m s$^{-1}$ amplitude at $z=0$ Mm (the amplitude of the vertical velocity at the bottom boundary is 0.3 m s$^{-1}$). As discussed previously, the sinusoidal acoustic wave introduced as a driver initially excites a broad wave spectrum, and the chromosphere is dominated by waves in the three-minute band during the first $\sim100$ min as a result of the resonant cavity. The use of a high amplitude driver in combination with the resonance generates highly non-linear waves with period in the 3 min band. These shocks are produced by the propagation of waves in the 3 min band (above the acoustic cutoff) from deeper layers generated by the driver during the initial stages of the simulation (see previous section). Their amplitude increases according to the drop of the density (this increase is higher than that of non-propagating evanescent waves) and develop into shocks at the chromosphere. They are not the result of the unrealistic merging of high-frequency waves in one-dimensional simulations \citep{Ulmschneider+etal2005}. Due to these nonlinearities, we have not been able to compute a simulation initialized with high amplitude and long enough to reach the stationary regime. Instead, this simulation was started with a low amplitude (similar to Simulation \ref*{sect:cavity_monochromatic}.1) and it was gradually increased. When the appropriate amplitude was reached for the five-minute wave, the driver no longer excites other frequencies, and the five-minute wave is purely monochromatic. This way, we have been able to explore the effect of a monochromatic wave with a five minutes period and realistic amplitude on a chromospheric cavity.

\begin{figure}[!ht] 
 \centering
 \includegraphics[width=9cm]{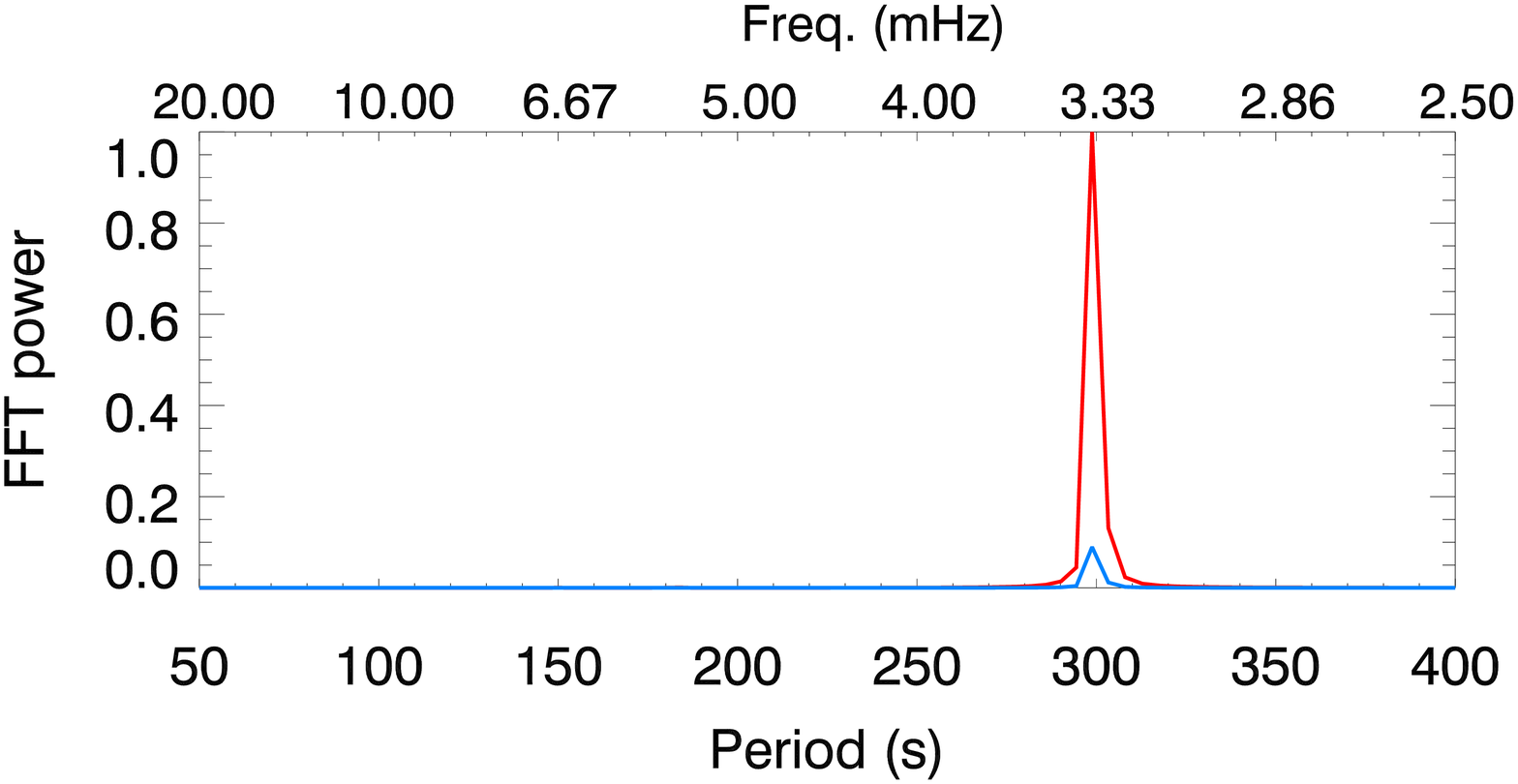}
  \caption{Power spectra of the vertical velocity from the stationary part of Simulation \ref*{sect:amplitude}.1 at $z=0.3$ Mm (blue line) and $z=1.4$ Mm (red line). Both spectra are normalized to the maximum value at $z=1.4$ Mm.}      
  \label{fig:monochromatic_cavity_amplitude}
\end{figure}

The duration of the simulation is 830 min. During the initial 400 min the amplitude of the bottom boundary driver is increasing, until it generates photospheric five-minute oscillations with 200 m s$^{-1}$ amplitude. Then, the amplitude of the driver stays constant, and the simulation reaches a stationary regime. Figure \ref{fig:monochromatic_cavity_amplitude} shows the power spectra of the stationary part of the simulation at the photosphere (blue line) and the chromosphere (red line). At both heights the power is concentrated at five minutes. These waves are evanescent, and their amplitude increases with height at a rate slower than that of propagating high-frequency waves. No trace of waves in the three-minute band is found. We can conclude that five-minute waves cannot excite oscillations in the three-minute band, even with the presence of a chromospheric cavity and realistic wave amplitudes. The excitation of standing waves in the chromospheric cavity requires the use of a broadband driver \citep{Botha+etal2011}. For simplicity, in the following we will focus on waves with low amplitude.

\subsection{Dependence of the resonance on the size of the cavity }
\label{sect:cavity_size}

\begin{table*}
\begin{center}
\caption[]{\label{table:cavity_size}
          {Summary of the numerical simulations performed for Sect. \ref{sect:cavity_size}}}
\begin{tabular}{lcccccc}
\hline\noalign{\smallskip}
Simulation \#			& 	Driver 		&	Height Cor.	& B 		&  Ampl		&$\tau_{\rm R}$ 	 &Power peak 		\\
				& 	 		&	 (Mm)		& (G)		&  (m s$^{-1}$)	& (s)			 &(s)		\\
\hline\noalign{\smallskip}	

\ref*{sect:cavity_size}.1	& 	Broadband	& 	1.8		& 0		&10$^{-3}$	&	$\infty$	&	175.6		\\
\ref*{sect:cavity_size}.2	& 	Broadband	& 	2.0		& 0		&10$^{-3}$	&	$\infty$	&	184.3		\\
\ref*{sect:cavity_size}.3	& 	Broadband	& 	2.2		& 0		&10$^{-3}$	&	$\infty$	&	192.9		\\
\ref*{sect:cavity_size}.4	& 	Broadband	& 	2.4		& 0		&10$^{-3}$	&	$\infty$	&	199.2		\\
\ref*{sect:cavity_size}.5	& 	Broadband	& 	2.6		& 0		&10$^{-3}$	&	$\infty$	&	204.9		\\
\ref*{sect:cavity_size}.6	& 	Broadband	& 	-		& 0		&10$^{-3}$	&	$\infty$	&	169.8		\\

 \hline 
  
\end{tabular}  
  
\begin{tablenotes}
\small
 \item {Same description as Table \ref{table:cavity_monochromatic}.}
\end{tablenotes} 
  
\end{center}
\end{table*}

In this section we aim to evaluate the frequency of the resonance for various atmospheric models. The steep temperature gradient of the transition region (which produces the reflection at the top of the chromospheric cavity) is shifted to different heights. We have introduced the broadband driver defined in Eq. \ref{eq:broadband}, since in the previous sections we showed that the excitation of the resonant mode requires the presence of that frequency in the driving mechanism.  

\begin{figure}[!ht] 
 \centering
 \includegraphics[width=9cm]{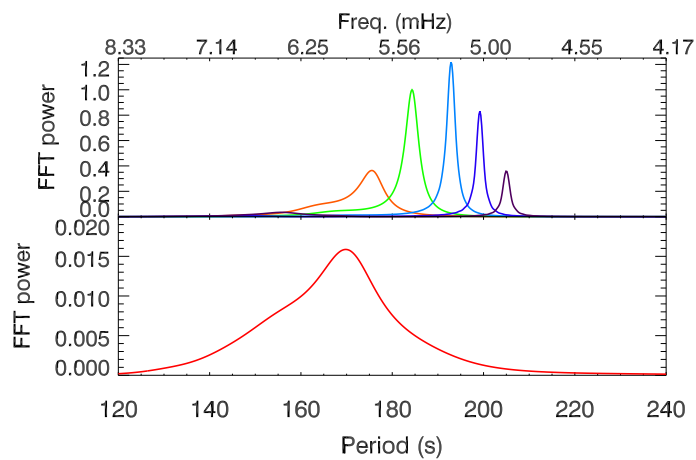}
  \caption{Top panel: Power spectra of the vertical velocity at $z=1.4$ Mm from numerical simulations with waves excited by a broadband driver and with the base of the corona at a different heights. The color code is the same from Fig. \ref{fig:models}, and the base of the corona is located at 1.8 Mm (orange, Simulation \ref*{sect:cavity_size}.1), 2.0 Mm (green, Simulation \ref*{sect:cavity_size}.2), 2.2 Mm (light blue, Simulation \ref*{sect:cavity_size}.3), 2.4 Mm (dark blue, Simulation \ref*{sect:cavity_size}.4), and 2.6 Mm (violet, Simulation \ref*{sect:cavity_size}.5). Bottom panel: Power spectra of the vertical velocity at $z=1.4$ Mm from  Simulation \ref*{sect:cavity_size}.6, using a broadband driver and without transition region and corona. All the power spectra are normalized to the maximum value of the power from Simulation \ref*{sect:cavity_size}.2 (height of the bottom of the corona at 2.0 Mm, non-magnetic, and without radiative losses).}      
  \label{fig:broadband_cavity_size}
\end{figure}

The top panel of Fig. \ref{fig:broadband_cavity_size} illustrates the power spectra of the chromospheric velocity for five different models (Simulations \ref*{sect:cavity_size}.1-\ref*{sect:cavity_size}.5). All of them show a sharp peak, but they differ in the period of the peak, the power, and the gradient of the spectra towards higher frequencies. The period of the resonance is approximately given by \citep[\eg,][]{Freij+etal2016}  

\begin{center}
\begin{equation}
\label{eq:resonance} 
P_{\rm r}\approx \frac{2L}{\bar{v}_cn},
\end{equation}
\end{center}

\noindent  where $L$ in the length of the cavity, $\bar{v}_c$ is the average speed of the wave (for acoustic waves, like this simulation, is the average of the sound speed $\bar{c}_s$), and $n$ is the harmonic number.  As expected, the longer the cavity the higher the period of the resonance. The measured power peak corresponds to the fundamental mode ($n=1$). For each simulation, assuming that the top of the cavity is given by the maximum of the acoustic cutoff ($z_{\rm 1}$, Fig. \ref{fig:models}), we can iterativelly determine $L=z_{\rm 1}-z_{\rm 0}$, where $z_{\rm 0}$ is the height of the bottom of the cavity. We start with an initial guess for $z_{\rm 0}$ and compute $\bar{c}_s$ between that height and $z_{\rm 1}$. This value of the average sound speed and the $P_{\rm r}$ obtained from the simulation are introduced in Eq. \ref{eq:resonance} in order to derive a new $z_{\rm 0}$. This process is repeated until convergence is found. The bottom of the cavity ranges from  $z=813$ to $z=1291$ km above the photosphere, being lower for the atmospheres with the transition region at a lower height. This is in agreement with the acoustic cutoff profiles. Models with the transition region at a lower height exhibit a resonance with higher frequency, and these waves can propagate to lower heights inside the cavity (Fig. \ref{fig:models}).

The relative power of the peaks obtained from each simulation depends on the duration of the simulation. Figure \ref{fig:broadband_cavity_size} shows the power spectra for the whole temporal series of 500 min. During the initial temporal steps the power of the cases with a lower transition region is stronger, but near the end of those simulations their waves have leaked the chromospheric cavity. For the models with higher transition region, the power inside the cavity last longer. For example, according to a wavelet analysis, the chromospheric power of the model with the highest transition region (Simulation \ref*{sect:cavity_size}.5) is reduced to a 10\% of its maximum value after 67 min, whereas for the model with the lowest transition region (Simulation \ref*{sect:cavity_size}.1) the same power reduction is found after 40 min. We speculate that escaping through the transition region is easier for the higher frequencies of the resonant mode of the atmospheres with lower transition region. In the case of the atmospheres with higher transition region, lower frequency waves are reflected more efficiently, and they remain longer inside the chromospheric cavity. The power spectra exhibit a tail with high power starting from the resonance power peak towards higher frequencies. This power enhancement is more prominent for the models with the transition region located at lower heights. This result is in agreement with the simulations from \citet{Snow+etal2015}.

As a reference, the bottom panel of Fig. \ref{fig:broadband_cavity_size} exhibits the power spectra obtained from a simulation with the same driver (with its maximum power at a period of five minutes) but without a chromospheric cavity formed by the presence of the transition region (Simulation \ref*{sect:cavity_size}.6). In this case, the power is also shifted towards the three-minute band. However, there are some significant differences. First, the power peak is located at around 6 mHz, just above the local maximum of the acoustic cutoff frequency. As discussed previously, this change in the dominant frequency is due to the linear propagation of waves, since the amplitude increase of waves above the acoustic cutoff is higher than that of evanescent waves. Second, the spectra is broader. There are no sharp peaks as a result of the resonance. And third, the power is almost two orders of magnitude lower than in some simulations with chromospheric cavity. In this simulation the high frequency waves travel through the domain and escape at the top boundary of the computational domain, so the wave power after the initial time steps is negligible. In addition, the resonance produces a remarkable enhancement of the power at a specific period, which makes the contrast even stronger.

\subsection{Magnetized atmospheres}
\label{sect:magnetic_field}

\begin{table*}
\begin{center}
\caption[]{\label{table:magnetic_field}
          {Summary of the numerical simulations performed for Sect. \ref{sect:magnetic_field}.}}
\begin{tabular}{lccccccc}
\hline\noalign{\smallskip}
Simulation \#			& 	Driver 		&	Height Cor.	& B 		&  Ampl		&$\tau_{\rm R}$ 	 &Power peak 		\\
				& 	 		&	 (Mm)		& (G)		&  (m s$^{-1}$)	& (s)			 &(s)			\\
\hline\noalign{\smallskip}	

\ref*{sect:magnetic_field}.1	& 	Broadband	& 	2.0		& 0		&10$^{-3}$	&	$\infty$	&	184.3		\\

\ref*{sect:magnetic_field}.2	& 	Broadband	& 	2.0		& 10		&10$^{-3}$	&	$\infty$	&	184.5		\\
\ref*{sect:magnetic_field}.3	& 	Broadband	& 	2.0		& 250		&10$^{-3}$	&	$\infty$	&	184.8		\\
\ref*{sect:magnetic_field}.4	& 	Broadband	& 	2.0		& 500		&10$^{-3}$	&	$\infty$	&	185.0		\\
\ref*{sect:magnetic_field}.5	& 	Broadband	& 	2.0		& 1000		&10$^{-3}$	&	$\infty$	&	185.2	\\
\ref*{sect:magnetic_field}.6	& 	Broadband	& 	2.0		& 1500		&10$^{-3}$	&	$\infty$	&	184.8		\\
\ref*{sect:magnetic_field}.7	& 	Broadband	& 	2.0		& 2000		&10$^{-3}$	&	$\infty$	&	184.5		\\
\ref*{sect:magnetic_field}.8	& 	Broadband	& 	2.0		& 2500		&10$^{-3}$	&	$\infty$	&	184.3		\\
\ref*{sect:magnetic_field}.9	& 	Broadband	& 	-		& 2500		&10$^{-3}$	&	$\infty$	&	179.5		\\
\ref*{sect:magnetic_field}.10	& 	Broadband	& 	-		& 0		&10$^{-3}$	&	$\infty$	&	169.8	 \\ 
 
\hline 
 
\end{tabular}  
  
\begin{tablenotes}
\small
 \item {Same description as Table \ref{table:cavity_monochromatic}.}
\end{tablenotes} 
  
\end{center}
\end{table*}

In this set of simulations (see Table \ref{table:magnetic_field}), the atmospheric model with the base of the corona located at $z=2.0$ Mm is permeated with an homogeneous vertical magnetic field. A different value of the magnetic field is chosen for each simulation. In order to avoid the small temporal step imposed by the extremely high coronal Alfv\'en speed, we have constrained the strength of the Lorentz force following \citet{Rempel+etal2009}. The Alfv\'en speed is artificially limited to \hbox{10,000 km s$^{-1}$}. At chromospheric layers and up to the transition region, the reduction of the Lorentz force is negligible. The whole chromospheric cavity is not affected by this Alfv\'en speed limiter. For one of the cases with strong magnetic field, we have performed another simulation with the Alfv\'en speed limited to \hbox{20,000 km s$^{-1}$} (instead of \hbox{10,000 km s$^{-1}$}), and we found no significant differences in the results.

The presence of the magnetic field changes the nature of the waves. Whereas in previous experiments, with the magnetic field set to zero, we have analyzed the behavior of acoustic waves, in this section we study the effect of the chromospheric cavity on new wave modes. The theoretical picture for an homogeneous atmosphere with constant magnetic field indicates that three wave modes are present: fast and slow magnetoacoustic waves and the Alfv\'en wave. In an inhomogenous atmosphere (like those analyzed in this work), the three wave modes are not completely decoupled, but still it is common to refer to these terms to discuss the waves. We are going to follow this approach. It is a good approximation in the regions where the Alfv\'en speed is much higher than the sound speed (or vice versa), but not so good around $\beta =c_s^2/v_A^2=1$. 

In our simulations, the driver is located in a high-$\beta$ region (where the sound speed is much higher than the Alfv\'en speed). The vertical velocity introduced by the driver generates a fast magnetoacoustic wave, whose behavior is similar to an acoustic wave. At the layer where the Alfv\'en speed approaches the sound speed (around the photosphere, depending on the magnetic field strength), the incident fast wave can be converted into a fast or a slow magnetoacoustic mode. In all the simulations discussed in the paper we have imposed a vertical magnetic field. Since the direction of propagation of the fast wave (in the high-$\beta$ region) is aligned with the direction of the magnetic field, conversion from fast to slow magnetoacoustic mode is favored at the height where $\beta\sim1$. The slow magnetoacoustic wave in the low-$\beta$ region behaves similar to an acoustic wave, but it propagates along field lines.

\begin{figure}[!ht] 
 \centering
 \includegraphics[width=9cm]{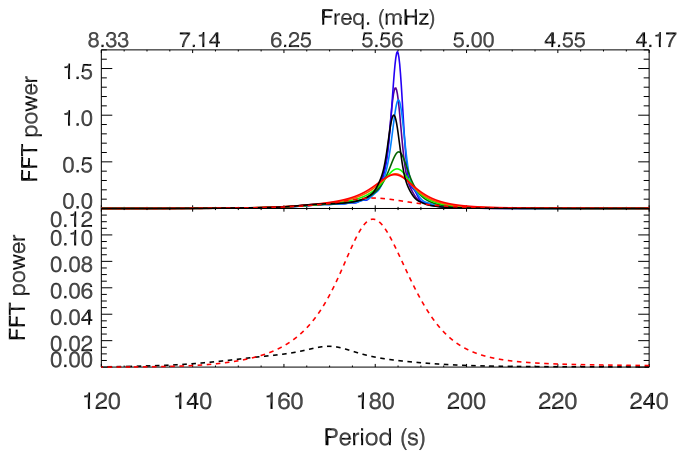}
  \caption{Power spectra of the vertical velocity at $z=1.4$ Mm from numerical simulations with waves excited by a broadband driver and various magnetic field strengths. The vertical magnetic field strength is 0 (black line, Simulation \ref*{sect:magnetic_field}.1), 10 G (violet line, Simulation \ref*{sect:magnetic_field}.2), 250 G (dark blue, Simulation \ref*{sect:magnetic_field}.3), 500 G (light blue, Simulation \ref*{sect:magnetic_field}.4), 1000 G (dark green, Simulation \ref*{sect:magnetic_field}.5), 1500 G (light green, Simulation \ref*{sect:magnetic_field}.6), 2000 G (orange, Simulation \ref*{sect:magnetic_field}.7), and 2500 G (red line, Simulation \ref*{sect:magnetic_field}.8). Solid lines indicate simulations with the base of the corona located at 2.0 Mm. Dashed lines indicate simulations without transition region and corona, with a vertical magnetic field strength of 0 (black line, bottom panel, \ref*{sect:magnetic_field}.10) and 2500 G (red lines, both panels, Simulation \ref*{sect:magnetic_field}.9). All the power spectra are normalized to the same value used for the normalization of the power spectra illustrated in Fig. \ref{fig:broadband_cavity_size}.}      
  \label{fig:broadband_cavity_B}
\end{figure}

Figure \ref{fig:broadband_cavity_B} shows the power spectra of the chromospheric velocity for various numerical simulations with different vertical magnetic field strengths, from 0 to 2500 G. The magnetic field produces small differences in the period of the power peak. For small values of the magnetic field strength (up to 1000 G), an increase in the field strength is associated to a increase in the period. For stronger magnetic field strengths, the period is reduced as higher values of the magnetic field strength are considered. According to Eq. \ref{eq:resonance}, the period of the fundamental mode of the resonance is determined by the size of the cavity and the wave speed. The wave speed of slow magnetoacoustic waves propagating along field lines in the low-$\beta$ region is the sound speed. However, part of the chromospheric cavity is located in a region where $\beta\sim 1$, especially in the cases with low magnetic field strength. The wave speed in that region will take some value between the sound speed and the Alfv\'en speed. This way, in the cases with magnetic field, $\bar{v}_c$ will be generally higher than in the field free cases. According to Eq. \ref{eq:resonance}, the increase of $P_{\rm r}$ with the magnetic field for low field strengths (up to 1000 G) must be due to an increase in the size of the cavity.  

The resonance in atmospheres with stronger magnetic field produces broader peaks with lower power. This indicates that the magnetic field enhances the leaking of the waves out of the chromospheric cavity. In order to test the capability of the waves to pass through the bottom boundary of the cavity (the local maximum in the magnetoacoustic cutoff frequency, Fig. \ref{fig:models}), we have compared two simulations without transition region and corona, one of them with a 2500 G magnetic field strength and the other field free. In these cases there is no transition region resonance, since waves escape through the top boundary. The chromospheric power is plotted in the bottom panel of Fig. \ref{fig:broadband_cavity_B}. In the simulation with $B=2500$ G the power is significantly higher, and the maximum power is located at a lower frequency. These results confirm that the magnetic field reduces the effective cutoff frequency, and low frequency waves can travel more easily through the temperature minimum when the magnetic field is present. As the magnetic field strength increases, the filtering of the lower frequencies in the spectrum is reduced. This is in qualitative agreement with the lower cutoff frequency derived for magnetoacoustic waves in an atmosphere with a vertical magnetic field (Fig. \ref{fig:models}). Thus, in strongly magnetized atmospheres, the chromospheric cavity behaves like a resonator with losses.

\subsection{Effect of radiative losses}
\label{sect:rad_trans}

\begin{table*}
\begin{center}
\caption[]{\label{table:rad_trans}
          {Summary of the numerical simulations performed for Sect. \ref{sect:rad_trans}.}}
\begin{tabular}{lcccccc}
\hline\noalign{\smallskip}
Simulation \#			& 	Driver 		&	Height Cor.	& B 		&  Ampl		&$\tau_{\rm R}$ 	 &Power peak 		\\
				& 	 		&	 (Mm)		& (G)		&  (m s$^{-1}$)	& (s)			 &(s)		\\
\hline\noalign{\smallskip}

\ref*{sect:rad_trans}.1		& 	Broadband	& 	2.0		& 0		&10$^{-3}$	&	$\infty$	&	184.3		\\

\ref*{sect:rad_trans}.2		& 	Broadband	& 	2.0		& 0		&10$^{-3}$	&	10		&	199.6	\\
\ref*{sect:rad_trans}.3		& 	Broadband	& 	2.0		& 0		&10$^{-3}$	&	20		&	198.2	\\
\ref*{sect:rad_trans}.4		& 	Broadband	& 	2.0		& 0		&10$^{-3}$	&	40		&	194.9	\\
\ref*{sect:rad_trans}.5		& 	Broadband	& 	2.0		& 0		&10$^{-3}$	&	70		&	188.3	\\
\ref*{sect:rad_trans}.6		& 	Broadband	& 	2.0		& 0		&10$^{-3}$	&	100		&	184.0	\\
\ref*{sect:rad_trans}.7		& 	Broadband	& 	2.0		& 0		&10$^{-3}$	&	180		&	181.2	\\
\ref*{sect:rad_trans}.8		& 	Broadband	& 	2.0		& 0		&10$^{-3}$	&	300		&	182.1	\\
\ref*{sect:rad_trans}.9		& 	Broadband	& 	2.0		& 0		&10$^{-3}$	&	600 		&	183.5	\\
\ref*{sect:rad_trans}.10		& 	Broadband	& 	2.0		& 0		&10$^{-3}$	&	1200		&	184.0	\\
\ref*{sect:rad_trans}.11		& 	Broadband	& 	2.0		& 0		&10$^{-3}$	&	3000 		&	184.2	\\
\ref*{sect:rad_trans}.12		& 	Broadband	& 	-		& 0		&10$^{-3}$	&	10 		&	199.2	\\
\ref*{sect:rad_trans}.13		& 	Broadband	& 	-		& 0		&10$^{-3}$	&	20 		&	197.2	\\
\ref*{sect:rad_trans}.14		& 	Broadband	& 	-		& 0		&10$^{-3}$	&	$\infty$	&	169.8	 \\ 
   
  \hline
  
\end{tabular}  
\begin{tablenotes}
\small
 \item {Same description as Table \ref{table:cavity_monochromatic}.}
\end{tablenotes} 
  
\end{center}
\end{table*}

We explore the impact of the radiative losses on the power spectra obtained from the resonant cavity. We have computed various simulations with differences in the characteristic radiative cooling time $\tau_{\rm R}$ (see Table \ref{table:rad_trans}). The radiative energy losses were implemented following Newton's cooling law:

\begin{equation}
\label{eq:newton_cooling}
Q_{\rm rad}=-c_v\frac{T_1}{\tau_R},
\end{equation}
\noindent where $T_1$ is the temperature perturbation and $c_v$ is the specific heat at constant volume. For each simulation we set a constant value for $\tau_{\rm R}$ above $z=0$ (in the simulations discussed in previous sections $\tau_{\rm R}\rightarrow \infty$). Radiative losses are neglected below $z=0$. In this section we made no effort to reproduce realistic values of the radiative losses in the actual umbral atmosphere. We aim to perform a parametric study on the effect of constant radiative losses on the chromospheric power spectra. 

\begin{figure}[!ht] 
 \centering
 \includegraphics[width=9cm]{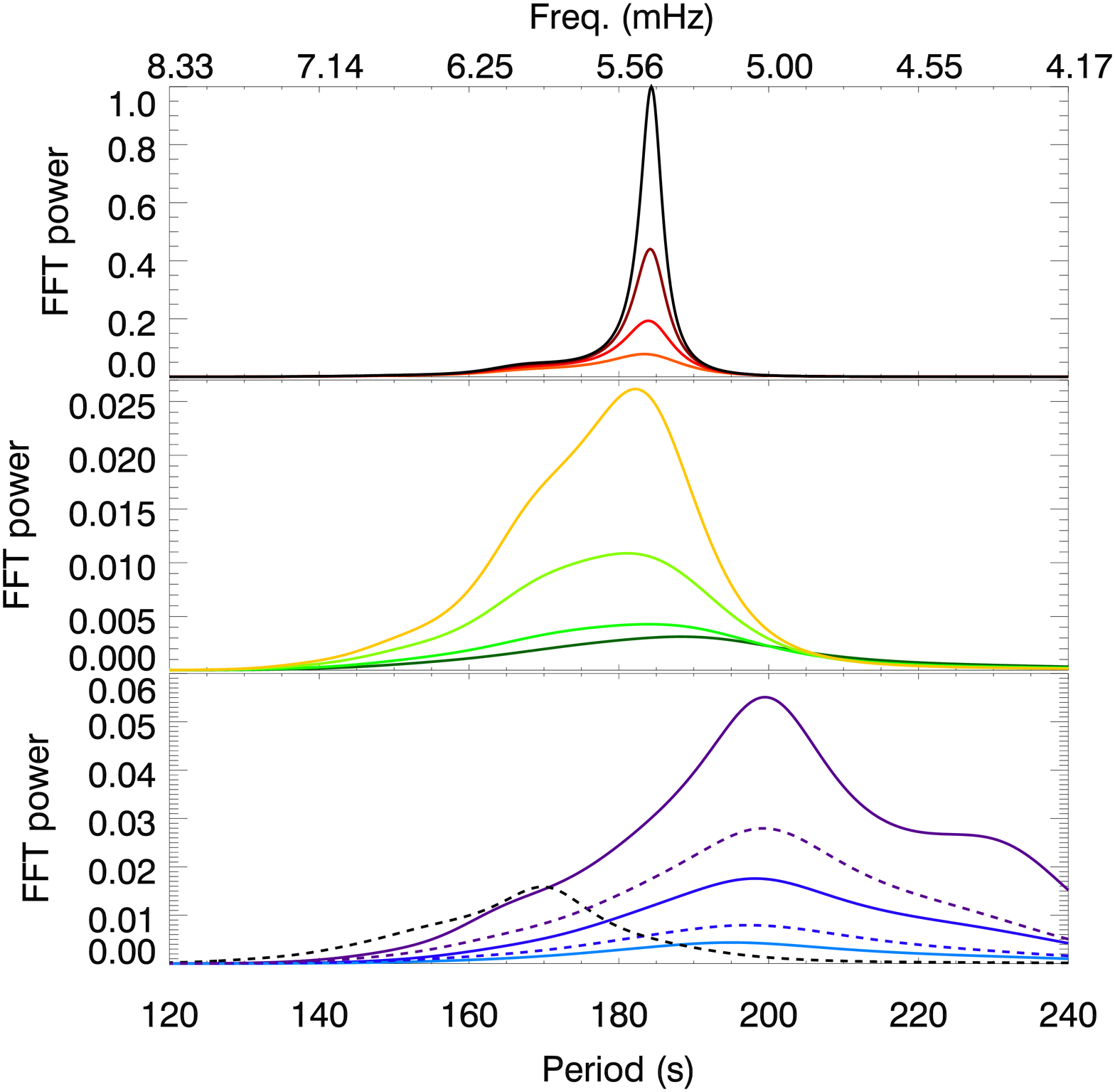}
  \caption{Power spectra of the vertical velocity at $z=1.4$ Mm from numerical simulations with various values of the radiative cooling time. In all the cases waves are excited by a broadband driver and the magnetic field is set to zero. Each panel illustrates a different set of simulations (note the differences in the range of the vertical axis). From top to bottom, the radiative cooling time is infinity (black line, Simulation \ref*{sect:rad_trans}.1), 3000 s (dark red line, Simulation \ref*{sect:rad_trans}.11), 1200 s (red line, Simulation \ref*{sect:rad_trans}.10), 600 s (orange line, Simulation \ref*{sect:rad_trans}.9), 300 s (yellow line, Simulation \ref*{sect:rad_trans}.8), 180 s (light green line, Simulation \ref*{sect:rad_trans}.7), 100 s (green line, Simulation \ref*{sect:rad_trans}.6), 70 s (dark green line, Simulation \ref*{sect:rad_trans}.5), 40 s (light blue line, Simulation \ref*{sect:rad_trans}.4), 20 s (dark blue line, Simulation \ref*{sect:rad_trans}.3), 10 s (violet line, Simulation \ref{sect:rad_trans}.2). Solid lines indicate simulations with the base of the corona located at 2.0 Mm. Dashed lines indicate simulations without transition region and corona, with a radiative cooling time of 10 s (violet dashed line, Simulation \ref*{sect:rad_trans}.12), 20 s (dark blue dashed line, Simulation \ref*{sect:rad_trans}.13), or infinity (black dashed line, Simulation \ref*{sect:rad_trans}.14). All the power spectra are normalized to the same value used for the normalization of the power spectra illustrated in Fig. \ref{fig:broadband_cavity_size}.}      
  \label{fig:broadband_cavity_radtrans}
\end{figure}

Figure \ref{fig:broadband_cavity_radtrans} illustrates the power spectra of the vertical velocity at $z=1.4$ Mm for all the cases with constant $\tau_{\rm R}$ above the photosphere. When the radiative losses are neglected (black solid line in top panel of Fig. \ref{fig:broadband_cavity_radtrans}), the power spectra exhibits a sharp peak with a period of 184.3 s, as discussed in previous sections. The introduction of low radiative losses (high values of $\tau_{\rm R}$) produces a striking reduction in the strength of the power peak and a broadening of the power distribution. With the presence of radiative losses there is no strict acoustic cutoff, and all frequencies are partially reflected and partially propagating \citep{Roberts1983, Centeno+etal2006, Khomenko+etal2008b}. The ratio of propagation versus reflection increases with the frequency. Similar to the effects of the magnetic field on the chromospheric cavity, the radiative losses also increase the losses of the resonator, and waves can escape from the cavity. 

The minimum power is found for $\tau_{\rm R}=40$ s. There are several aspects that determine the power of the waves at chromospheric heights. First, lower radiative losses produce a better resonant cavity, and the power at the resonant frequency (around 184 s) is enhanced more efficiently. Second, higher radiative losses increase the damping of the waves, since they lose energy as they propagate from the photosphere to the chromosphere. And third, higher radiative losses reduce the effective acoustic cutoff frequency, allowing more wave energy to reach the chromosphere. This explain the increase of the maximum power in the cases with $\tau_{\rm R}<40$ s (dark blue and violet lines in bottom panel of Fig. \ref{fig:broadband_cavity_radtrans}). Thanks to the reduction of the acoustic cutoff, waves with a period around 200 s can propagate from the photosphere to the chromosphere. Since the broadband driver introduces more wave energy at this period that at shorter periods, the chromospheric power of the cases with very low $\tau_{\rm R}$ is stronger despite the enhanced radiative losses. 

The power spectra as a function of the radiative cooling time exhibits several regimes. For the cases with longer radiative cooling time (above $\sim 180$ s), it is dominated by the resonant mode, with a narrow band at 184 s and a tail with high power towards shorter periods. The strength of the resonant peak decreases with shorter radiative cooling times. For those cases with very short radiative cooling times, the chromospheric spectra is not dominated by the resonance, but by the power of waves propagating from deeper layers up to the chromosphere, instead. This produces a shift of the power to longer periods, since the acoustic cutoff frequency is reduced due to the radiative losses. 

In order to confirm this interpretation, we have analyzed some cases without transition region and corona. The dashed lines in the bottom panel of Fig. \ref{fig:broadband_cavity_radtrans} show the power spectra for two simulations without transition region and short radiative cooling times (dark blue and violet dashed lines, Simulations \ref*{sect:rad_trans}.12 and \ref*{sect:rad_trans}.13). The power distribution is similar to that obtained for the simulations with transition region (dark blue and violet solid lines). However, when the transition region is present the power is stronger, since the sharp temperature gradient at the transition region acts as a barrier and partially reflects the waves. This way, more wave energy remains at the chromosphere. The dashed black line (bottom panel of Fig. \ref{fig:broadband_cavity_radtrans}) shows the power spectra from a simulation without radiative losses and without transition region (Simulation \ref*{sect:rad_trans}.14). In this case the chromospheric power is concentrated at shorter periods than in the cases with radiative losses (dark blue and violet dashed lines), because in the later the effective acoustic cutoff frequencies are reduced.

\subsection{Chromospheric cavity with radiative losses and magnetic field}
\label{sect:spiegel}

\begin{table*}
\begin{center}
\caption[]{\label{table:spiegel}
          {Summary of the numerical simulations performed for Sect. \ref{sect:spiegel}.}}
\begin{tabular}{lcccccc}
\hline\noalign{\smallskip}
Simulation \#			& 	Driver 		&	Height Cor.	& B 		&  Ampl		&$\tau_{\rm R}$ 	 &Power peak 	\\
				& 	 		&	 (Mm)		& (G)		&  (m s$^{-1}$)	& (s)			 &(s)			\\
\hline\noalign{\smallskip}	
\ref*{sect:spiegel}.1	& 	Broadband	& 	2.0		& 0		&10$^{-3}$	&	Spiegel		&	183.8	\\
\ref*{sect:spiegel}.2	& 	Broadband	& 	2.0		& 2500		&10$^{-3}$	&	Spiegel		&	186.0	\\
\ref*{sect:spiegel}.3	& 	Broadband	& 	-		& 0		&10$^{-3}$	&	Spiegel		&	171.5	\\
\ref*{sect:spiegel}.4	& 	Broadband	& 	-		& 2500		&10$^{-3}$	&	Spiegel		&	182.6	\\

\hline

\end{tabular}

\begin{tablenotes}
\small
 \item {Same description as Table \ref{table:cavity_monochromatic}. The term ``Spiegel'' in the sixth column indicates that the value of the radiative cooling time in the photosphere and low chromosphere is computed using Eq. \ref{eq:spiegel}, while radiative losses are neglected in the rest of the domain.}
\end{tablenotes} 
  
\end{center}
\end{table*}

\begin{figure}[!ht] 
 \centering
 \includegraphics[width=9cm]{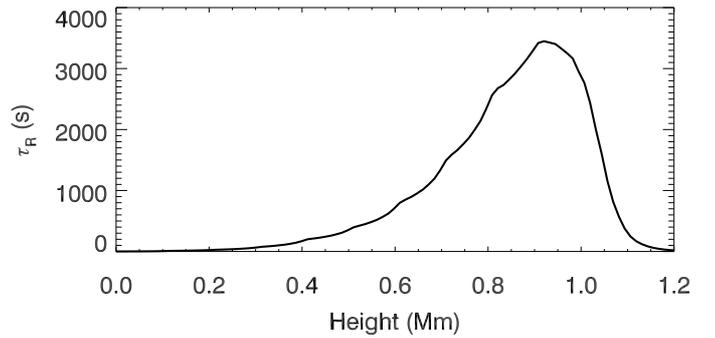}
  \caption{Radiative cooling time in the model atmosphere according to \citet{Spiegel1957}}      
  \label{fig:spiegel}
\end{figure}

Following \citet{Spiegel1957}, the radiative cooling time can be obtained as 

\begin{equation}
\label{eq:spiegel}
\tau_{\rm R}=\frac{\rho c_v}{16\chi\sigma_R T^3},
\end{equation}

\noindent where $T$ is the temperature, $\chi$ is the gray absorption coefficient, and $\sigma_R$ is the Stefan-Boltzmann constant. This expression is defined in the optically thin limit and assuming local thermodynamic equilibrium and, thus, it is only applicable at photospheric heights. Figure \ref{fig:spiegel} shows the vertical stratification of $\tau_R$ in our model atmosphere. It takes values of the order of seconds at the photosphere and rapidly increases with height, showing values around $\tau_R\sim3500$ s at $z=0.9$ Mm. At higher layers $\tau_R$ decreases with height, with $\tau_R=85$ s at $z=1.15$ Mm (where the temperature of the atmospheric model is 6360 K). This radiative cooling time is in agreement with the chromospheric value determined by \citet{Giovanelli1978}. Above that layer, \citet{Giovanelli1978} estimated an increase of $\tau_R$ up to 390 s for $T=8400$ K, whereas according to \citet{Spiegel1957} formula the radiative cooling time keeps decreasing. For simplicity, the radiative loss function is only applied in the region between $z=0.2$ and $z=1.15$ Mm. Out of this region the applicability of Eq. \ref{eq:spiegel} is uncertain. Assuming adiabatic propagation is justified for chromospheric layers, those with low optical depth \citep{Schmieder1977}. Below $z=0.2$ Mm, Spiegel's formula exhibits unrealistic low values for the radiative cooling time due to the optically thin treatment. We have followed the same approach from \citet{Ulmschneider1971}, assuming a completely optically thick atmosphere below a certain arbitrary height. \citet{Ulmschneider1971} chose a height between $z=0.11$ and $z=0.14$ Mm as the lower limit of applicability of Eq. \ref{eq:spiegel}. We set it a $z=0.2$ Mm. At at height, the minimum value of $\tau_R$ used in these simulations is found ($\tau_R=24$ s).

\begin{figure}[!ht] 
 \centering
 \includegraphics[width=9cm]{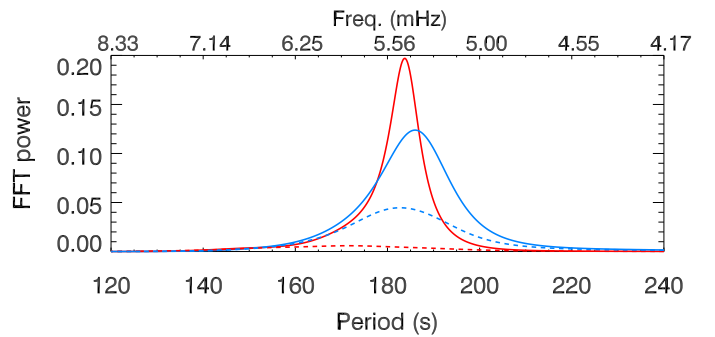}
  \caption{Power spectra of the vertical velocity at $z=1.4$ Mm from numerical simulations with the photospheric radiative losses estimated from \citet{Spiegel1957}. Red lines correspond to cases with the magnetic field set to zero, and blue lines to a magnetic field strength of 2500 G. Solid lines represent the cases with the base of the corona located at $z=2.0$ Mm, and dashed lines to cases without corona and transition region. All the power spectra are normalized to the same value used for the normalization of the power spectra illustrated in Fig. \ref{fig:broadband_cavity_size}.}     
  \label{fig:broadband_cavity_spiegel}
\end{figure}

Figure \ref{fig:broadband_cavity_spiegel} illustrates the chromospheric power spectra measured from simulations with the radiative losses implemented following Eq. \ref{eq:spiegel} (see Table \ref{table:spiegel}). We compare non-magnetized (red lines) and magnetized atmospheres (2500 G, blue lines) for the simulations with chromospheric cavity (solid line) and without transition region (dashed lines). A comparison of the two simulations with the magnetic field set to zero shows remarkable differences. When the waves are reflected at the top of the chromosphere by the steep temperature gradient (solid line), a resonance is form with the power peak at a period of 183.8 s. On the contrary, if waves are able to escape through the top boundary, the chromospheric power is significantly lower, since it is only due to the linear propagation of high-frequency waves from deeper layers. In this case, the chromospheric power spectra is dominated by waves with frequency above the effective magnetoacoustic cutoff frequency (taking into account the effect of the radiative losses) and exhibits a maximum value at a period of 171.5 s.

In the case of the simulations with the atmosphere permeated by a vertical magnetic field, the maximum of the power is obtained at 186.0 and 182.6 s for the simulations with and without transition region, respectively. Significant differences are found in the power between both cases, although these differences are not as striking as in the unmagnetized cases. The combined action of radiative losses and magnetic field strength facilitates the leaking of the waves, and the resonance produces a power spectra with lower power and broader. With realistic photospheric radiative losses and realistic values of the umbral magnetic field, the resonance is less efficient.

\section{Discussion}
\label{sect:discussion}

In this paper, we study the chromospheric power spectra from numerical simulations of wave propagation in a sunspot umbra. We explore the response of the umbral atmosphere to wave excitation in cases with different parameters. Our numerical experiments include changes in the driving mechanism, the height of the transition region and corona, the vertical magnetic field strength, and the radiative losses. 

One of the basic approaches of this paper is the comparison of the power spectra in simulations with a chromospheric cavity (when the transition region and corona are present) and without a resonant cavity. In the later, waves escape through the top boundary of the computational domain, which is located at the upper chromosphere. These two numerical setups allow us to evaluate two mechanisms proposed to explain the three-minute oscillations observed at chromospheric heights. Firstly, the action of a chromospheric resonant cavity \citep{Zhugzhda+Locans1981, Zhugzhda+etal1983,Gurman+Leibacher1984}. Between the temperature gradient of the photosphere and that at the transition region, the acoustic (and magnetoacoustic) cutoff frequency exhibits a local minimum. Slow magnetoacoustic waves in a certain frequency range are partially trapped inside this region. The resonant modes are enhanced, while the frequencies outside the resonance range slowly leak. Secondly, the linear propagation of waves with frequencies above the cutoff frequency. Their amplitude increases due to the drop of the density, and at the chromosphere they dominate over the evanescent low frequency waves \citep{Fleck+Schmitz1991, Centeno+etal2006}.  

Our simulations show that a monochromatic wave with a five minute period traveling from the interior to the corona cannot generate chromospheric oscillations in the three-minute band, as those measured in sunspot observations. This finding contrasts with the recent conclusions from \citet{Kraskiewicz+etal2019} (in ideal MHD) and \citet{Wojcik+etal2018} (using a two-fluid model), who reported that the energy from a driving period of five minutes can be transferred to shorter periods as a result of wave reflection. Our simulations reproduce the same effect reported by \citet{Kraskiewicz+etal2019} and \citet{Wojcik+etal2018}. However, chromospheric three-minute waves are only found if the driver excites these frequencies. In our simulations, introducing as a driver at the bottom boundary a sinusoidal acoustic wave (including the perturbations in velocity, pressure, and density with the appropriate amplitude and phase relations), we have proven that the oscillations in the three-minute band are only excited during the initial stages of the simulation. We interpret that they are the result of the small power introduced in the three-minute band by the driver with five minutes period from the head of the wave during the transient at the initial times of the simulations. These three-minute waves are free to propagate upward, increasing their amplitude, and they initially dominate the chromospheric power spectra due to the spatial attenuation of the evanescent low frequency waves (five-minute band), until they escape from the computational domain. If the transition region is present, a chromospheric cavity is formed, and the power of the three-minute waves is greatly enhanced by the resonance. This power is progressively reduced as the waves leak the chromospheric cavity. When the stationary stage of the simulation is reached (after 230 minutes in the simulation using an acoustic wave as driver), we obtain a pure monochromatic driver and five-minute oscillations in all the domain. 

One fundamental difference of our numerical set up with those from \citet{Kraskiewicz+etal2019} and \citet{Wojcik+etal2018} is the moving piston introduced at the bottom boundary. They used a sinusoidal wave in the vertical velocity, without including the perturbations in the rest of variables with the amplitude and phase relations for an acoustic wave. As discussed by \citet{Fleck+Schmitz1991}, oscillations at the three-minute band are already present in this driver due to the relations between the thermodynamic quantities. As a sanity check, we repeated the simulations using as a driver a sinusoidal wave only in the vertical velocity. Those simulations require a much longer time to reach a stationary regime, and the chromospheric three-minute oscillations dominate for a long time. We can conclude that both mechanisms discussed in this paper for generating the three-minute power observed at the chromosphere (linear propagation of high frequency waves and chromospheric resonant cavity) require the excitation of three-minute waves at lower heights.

For the first time, we have explored the impact on the resonant cavity of two key ingredients for the modeling of umbral atmospheres, such as the magnetic field and the radiative losses. Up to now, the radiative losses had been neglected by analytical and numerical works. With regards to the magnetic field, it was assumed that its only role was to serve as a waveguide for slow magnetoacoustic waves, and low magnetic field strengths were chosen \citep[$\sim10$ G,][]{Botha+etal2011, Snow+etal2015}. However, magnetic field is known to modify the cutoff frequency of magnetoacoustic waves \citep[\eg,][]{Thomas1982, Thomas1983, Campos1986, Stark+Musielak1993, Roberts2006}. Umbral atmospheres with realistic chromospheric magnetic field strength values \citep[$\sim2000$ G, \eg,][]{delaCruz-Rodriguez+etal2013} exhibit notable differences in the permeability of the bottom boundary of the resonant cavity. The power spectra of the resonant mode produced by magnetized atmospheres shows slight variations in the frequency of the power peak, and striking differences in the strength and broadening. The magnetic field enhances the wave leakage through the photospheric boundary of the resonant cavity, and it behaves like a resonator with significant losses. In this work we have focused on sunspot umbrae and we have only analyzed vertical magnetic fields, but the magnetoacoustic cutoff frequency can also be lowered by inclined magnetic fields \citep{Bel+Leroy1977, Jefferies+etal2006}.

Radiative energy losses also produce a reduction of the cutoff frequency \citep{Roberts1983, Centeno+etal2006, Khomenko+etal2008b}. Their effect on the resonance is similar to that of the magnetic field, since they contribute to the leaking of the waves out of the resonant cavity. The changes in the cutoff frequencies are also evident in the simulations without resonant cavity produced by the transition region. When the radiative losses and/or the magnetic field are present, the wave power that reaches the chromosphere is shifted to lower frequencies (longer periods).

\section{Conclusions}
\label{sect:conclusions}

The numerical investigation presented in this paper shows that two of the mechanisms proposed to explain the chromospheric three-minute oscillations (linear propagation of high frequency waves and chromospheric resonant cavity) can actually reproduce the observed shift in the period of the wave power with height. We conclude that both mechanisms require the excitation of three-minute oscillations at deeper layers. Since waves are driven below the photosphere by turbulent plasma motions produced by magnetoconvection, it is expected that broadband excitation will always be present and, thus, both mechanisms can take place at the same time. The question remains of what is the relative importance of each of them. We have proven that the radiative losses and the magnetic field favor the leaking of the waves through the boundaries of the resonant cavity, and they reduce the enhancement of the wave power due to the resonance. The simulations using realistic values for the magnetic field strength and the photospheric radiative relaxation time demonstrate that chromospheric three-minute oscillations are obtained without a chromospheric cavity, but when it is present their power is enhanced almost a factor of three.

\begin{acknowledgements} 
Financial support from the State Research Agency (AEI) of the Spanish Ministry of Science, Innovation and Universities (MCIU) and the European Regional Development Fund (FEDER) under grant with reference PGC2018-097611-A-I00 is gratefully acknowledged. Wavelet software was provided by C. Torrence and G. Compo and is available at URL: http://paos.colorado.edu/research/wavelets/. The author wishes to acknowledge the contribution of Teide High-Performance Computing facilities to the results of this research. TeideHPC facilities are provided by the Instituto Tecnol\'ogico y de \hbox{Energ\'ias} Renovables (ITER, SA). URL: http://teidehpc.iter.es. 
\end{acknowledgements}

\bibliographystyle{aa} 
\bibliography{biblio.bib}

\end{document}